\documentclass[aps,pra,twocolumn,showpacs,epsfig,floats]{revtex4-1}
\usepackage{amsmath}
\usepackage{color}
\usepackage{graphicx}
\usepackage{epsfig}
\usepackage{amsfonts}
\usepackage{mathrsfs}
\usepackage{amsmath}
\usepackage{float}

\begin{document} 

\title{Dissipative Bose-Josephson junction coupled to bosonic baths}
 
\author{Sudip Sinha and S. Sinha}
\affiliation{Indian Institute of Science Education and
Research-Kolkata, Mohanpur, Nadia-741246, India}
 
\date{\today}

\begin{abstract}
We investigate the effect of dissipation in a Bose-Josephson junction (BJJ) coupled to bath of bosons at two sites. 
Apart from the dynamical transition due to repulsive interactions, the BJJ undergoes a quantum phase transition by increasing the coupling strength with the bath modes. We analyze this system by mapping to an equivalent spin model coupled to the bosonic modes. The excitation energies and fluctuation of number imbalance are obtained within Holstein-Primakoff approximation, which exhibit vanishing of energy gap and enhanced quantum fluctuations at the critical coupling. We study the dynamics of BJJ using time dependent variational method and analyze stability of different types of steady states. As a special case we study in details the phase space dynamics of BJJ coupled to a single mode, which reveals diffusive and incoherent behaviour with increasing coupling to the bath mode. The dynamical steady states corresponding to the Pi-oscillation and self-trapped state become unstable in the region where their oscillation frequencies are in resonance with the bath modes.
We investigate the time evolution of number imbalance and relative phase in presence of Ohmic bath with Gaussian noise to incorporate thermal fluctuations. Apart from damping of Josephson oscillations and transition to symmetry broken state for strong coupling we observe decay of Pi-oscillation and self-trapped state to the ground state as a result of dissipation. Variation of phase fluctuation with temperature of the bath shows similar behaviour as observed in experiment.
Finally we discuss the experimental setup to study the observable effects of dissipation in BJJ.   
\end{abstract}


\maketitle

\section{Introduction} 
Quantum dissipation is well studied subject in condensed matter physics which describes various interesting effects in different quantum systems arising from its coupling to the environment\cite{weiss}. In recent years, the study of quantum dissipation in ultracold quantum gases attracts much attention due to the presence of various loss processes affecting the coherence properties of the matter wave and to investigate dynamics of such open quantum system\cite{bec_diss}.
Also ultracold atomic system has become a testbed to study interplay between interaction and dissipation in a quantum many body system out of equilibrium \cite{bec_bath,driven_dissipative}. Apart from dephasing and relaxation dynamics, the combined effect of interaction and dissipation in open quantum systems can give rise to non-equilibrium steady states and transition between them\cite{dynamic_transition,ryd}.
Bose Josephson junction (BJJ) is an ideal system where both the effects of interaction and dissipation can be studied from the phase coherent collective dynamics\cite{leggett}. In experiments, Bose Josephson junction can be created by coherently coupling two Bose-Einstein condensates (BEC) in double well trap\cite{BJJ_exp,BJJ_expst}. Due to the inter particle interaction, the BJJ exhibits different types of oscillations\cite{BJJ_osc} and dynamical transition between them, which has been observed experimentally\cite{BJJ_expbifurcation}. Most remarkable phenomena is the appearance of `self-trapped' state above a critical strength of repulsive interaction where the symmetry between two condensates of the BJJ is broken and a number imbalance between them is spontaneously generated \cite{BJJ_expst}. Unlike quantum phase transition of the ground state, this `self-trapped' state represents a dynamically stable broken symmetry steady state arising due to repulsive inter particle interaction\cite{BJJ_osc,dyntransition}. Dissipation originating from finite temperature effects, coupling to quasi-particles and interaction with environment can lead to dephasing and decoherence in BJJ. Diffusion of relative phase plays a crucial role in controlling coherence in Josephson dynamics. Phase diffusion and heating effect in a BJJ due to thermal fluctuations has already been observed in experiment\cite{exp_phase}. 
To understand dephasing in BJJ, thermal and quantum fluctuations\cite{stringari_thermphase} as well as other dissipative processes have been studied within phenomenological models which gives rise to more complex dynamical behaviour\cite{walls1,vardi_phasenoise,wimberger1,minguzzi_phase,vardi1}. 
Also the dissipative dynamics of BJJ in the presence of quasi-particle excitations has been studied theoretically\cite{kroha1,kroha2}.
Josephson oscillation in presence of phenomenological damping term exhibits interesting non-linear dynamics\cite{BJJ_oscdamp}.
 
Following the seminal work of Caldeira and Leggett\cite{cl}, dissipation can be treated at the microscopic level by coupling the quantum system to external reservoir. 
In a similar manner, dissipation in two level system is introduced in well known `spin-boson' model exhibiting rich dynamical behaviour and quantum phase transition\cite{spinboson1,weiss}. In the experiments on ultracold atomic system, the thermally excited atoms constitute a heat bath in a natural way giving rise to dissipation. There is proposal to realize `spin-boson' model by coupling a micro-trap with the bath of phonon excitations of an elongated condensate\cite{sb_bec}. Effect of quantum dissipation on tunneling rate has also been observed experimentally\cite{diss_tunelexp}.
Moreover, impurity models can be realized in ultracold atomic system by coupling the impurity atom with the phonon bath of condensate\cite{imp_bec}. Due to the tunability in cold atomic systems, it is possible to engineer the bath modes and their spectral properties. Presence of thermal atoms and phonon baths are natural source of dissipation to the dynamics of BJJ formed by two weakly coupled condensates. Moreover, by coupling the condensates with finite cavity modes opens up the possibility to study more complex dynamics\cite{cavity}. In this work we follow the Caldeira-Leggett like approach to incorporate dissipation in BJJ by coupling it to heat baths. The ground state, excitations and fate of various oscillatory modes in presence of dissipation are main focus of the present study. 

Within tight binding approximation, the Bose-Josephson junction is simply described by a two site Bose-Hubbard model(BHM) with on-site interactions\cite{walls}. To study the dissipative effects in BJJ, in this work we consider a two-site BHM coupled to two reservoirs described by the Hamiltonian,
\begin{eqnarray}
H &=&-J(a_{L}^{\dagger}a_{R} + a_{R}^{\dagger}a_{L}) + \frac{U}{N}\left[\hat{n}_{L}(\hat{n}_{L}-1) + \hat{n}_{R}(\hat{n}_{R} -1)\right] \nonumber\\
&+& H_{IL} + H_{IR} + H_{BL} + H_{BR}.
\label{bhm_bath}
\end{eqnarray}
The total Hamiltonian $H$ consists of three parts, first part describes a two site BHM where the sites are denoted by L(left) and R(right), $a_{L/R}$, $\hat{n}_{L/R}$ are boson annihilation and number operators in respective sites, $J$ is the hopping amplitude between two sites and $U$ is on-site interaction strength. Total number of bosons in BJJ is $N$, and the interaction strength $U$ is scaled by the number of particles in order to maintain the extensivity of energy.
Other parts of $H$ describe the bosonic baths and their coupling with two different sites of the BHM. Here we consider simple non-interacting bosonic baths with linear coupling to density operators at respective sites described by the Hamiltonians,
\begin{eqnarray}
H_{Bi} & = & \sum_{k}\hbar \omega_{k} b_{ik}^{\dagger}b_{ik},\\
H_{Ii} & = & \sum_{k} \frac{\lambda_{ik}}{\sqrt{N}}\hat{n}_{i}(b_{ik}^{\dagger} + b_{ik})
\label{H_bath}
\end{eqnarray}
where the site index is $i=L,R$, $b_{ik}$ is bosonic annihilation operator of $i$th bath with momentum $k$ and energy $\hbar \omega_{k}$. Above Hamiltonian describes a very general model of dissipative two site BHM where total number of bosons $N$ in two sites remains conserved. For different values of $N$, this model is equivalent to a generalized spin-boson model with spin $N/2$ coupled to bosonic bath. For $N=1$, above Hamiltonian reduces to well known spin boson model\cite{spinboson1} where spin up(down) state corresponds to single boson in L(R) site.
For sufficiently large number of particles $N\gg 1$, the two site BHM describes a BJJ with phase coherent dynamics. However the influence of bath on phase coherence and oscillation modes of BJJ are relevant issues which are addressed in the present work.

This paper is organized as follows. In section II we analyze the model semi-classically for $N\gg 1$ to obtain ground state, collective excitations and quantum fluctuations which reveals quantum phase transition in BJJ due to dissipation. The dynamics of BJJ within time dependent variational approach and the steady states and their stability in presence of bath modes are presented in section III. Next in section IV, we study the dynamics of BJJ coupled to a single bosonic mode. Dissipative dynamics and thermal fluctuations in the presence of Ohmic bath are discussed in section V. Finally in section VI, we summarize the results and discuss possible experiments to observe the effects of dissipation.
 
\section{Quantum phase transition in dissipative BJJ} 
To analyze the ground state and excitations of the dissipative BJJ coupled to the heat bath, we simplify the model given in Eq.\ref{bhm_bath}.  
Two site Bose-Hubbard model with fixed number of bosons can be written as a spin Hamiltonian using the Schwinger boson representation,
$\hat{S}_{x}=a_{L}^{\dagger}a_{R} + a_{R}^{\dagger}a_{L}$ and $\hat{S}_{z}= \hat{n}_{L}-\hat{n}_{R}$, with magnitude of total spin $S=N/2$. 
Resulting spin Hamiltonian $H_{BJJ} = -J\hat{S}_x+\frac{U}{2S}\hat{S}_{z}^{2}$ with anti-ferromagnetic interaction for $U>0$ describes the BJJ within tight binding approximation\cite{walls}. Various interesting properties of this Hamiltonian related to many-body states, semi-classical spectrum and quantum dynamics have been explored theoretically\cite{anglin,vardi1,minguzzi_mott,carr,trippenbach,julia_diaz,vardi2,sinha}.
To simplify the model further, we consider a linear shift of the bath modes $\tilde{b}_{ik}=b_{ik} -\frac{\sqrt{2S}\lambda_{ik}}{2\hbar\omega_{k}}$ followed by an unitary transformation of the bosonic operators, 
\begin{eqnarray}
b_{k} & = & \frac{\lambda_{Rk}}{\lambda_k}b_{Rk} -\frac{\lambda_{Lk}}{\lambda_k}b_{Lk}\nonumber\\
B_{k} & = & \frac{\lambda_{Lk}}{\lambda_k}b_{Rk} +\frac{\lambda_{Rk}}{\lambda_k}b_{Lk} 
\label{new_b}
\end{eqnarray}
where $\lambda_k = \sqrt{\lambda_{Lk}^{2} + \lambda_{Rk}^{2}}$. Under this transformation, one of the bath modes $B_{k}$ decouple and the original Hamiltonian (Eq.\ref{bhm_bath}) of two site BHM in presence of two heat baths reduces to a spin system coupled to one bosonic bath described by the Hamiltonian,
\begin{eqnarray}
 H=&&-J\hat{S}_x+\frac{U}{2S}\hat{S}_{z}^{2} -\frac{\gamma_{-}}{2}\hat{S}_z \nonumber\\ &&+\frac{\hat{S}_{z}}{2\sqrt{N}}\sum_k\lambda_{k}(b_{k}+ b_{k}^\dagger)+\sum_k\hbar\omega_{k}b_{k}^\dagger b_{k},
\label{spin_ham}
\end{eqnarray}
where we define important parameters of the system arising from the coupling with two heat baths,
\begin{equation}
\gamma_{\pm} = \sum_k(\lambda_{Rk}^{2}\pm \lambda_{Lk}^2)/\hbar\omega_{k}.
\label{gamma}
\end{equation}
In above Hamiltonian, a new term $\frac{\gamma_{-}}{2}\hat{S}_z$ related to number imbalance in two sites is generated due to unequal coupling to two bosonic baths. In rest of the paper, we scale all energies by $J$ and time by $\hbar/J$ by setting $J=1$.

\begin{figure}[ht]
\centering
\includegraphics[height=7cm,width=8cm]{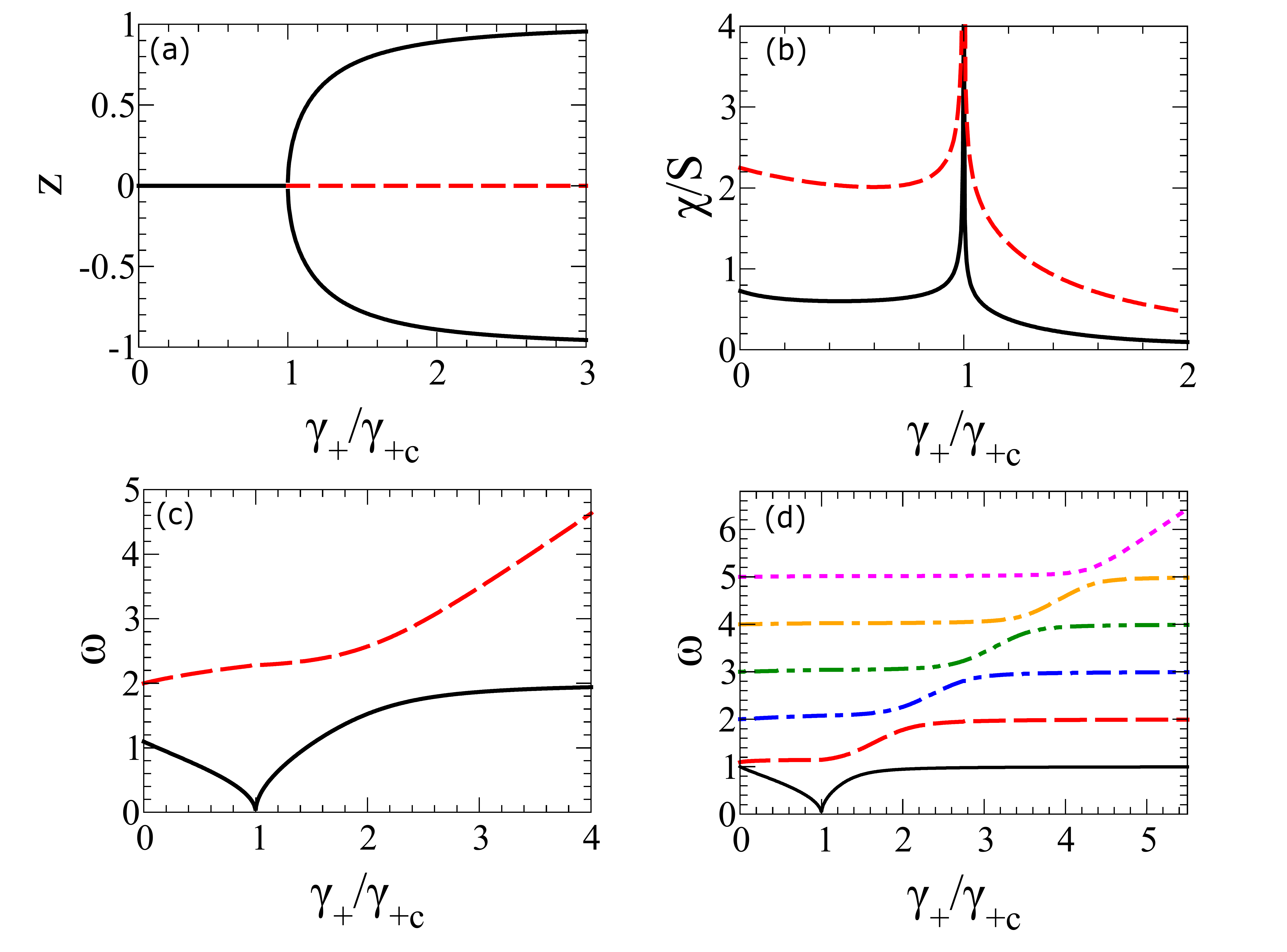}
\caption{Quantum phase transition by tuning dissipative coupling $\gamma_{+}$ for fixed interaction strength $U=0.2$. (a) Bifurcation diagram of imbalance fraction z showing quantum phase transition. (b) Spin fluctuation for single bath mode (solid line) and multi-modes (dashed line) near the critical point. Excitation energies (c) for single bath mode with $\omega_0 = 2$ and (d) for five modes where frequencies and coupling constants are chosen according to Eq.\ref{discretization} with $\Omega=5$. Vanishing of energy gap confirms quantum phase transition. All energies are in units of J.}
\label{fig:1}
\end{figure}

For $S=N/2 \gg 1$, we analyze the spin model in Eq.\ref{spin_ham} semi-classically where $1/S$ becomes a small parameter. We consider a rotation of coordinate in the $x-z$ plane at an angle $\theta$ with the $z$ axis so that the classical spin vector $\vec{\tilde{S}}$ in the rotated frame is aligned along the $z$-axis. Under this rotation, the components of the spin are related by,
\begin{eqnarray}
\hat{S}_z &=& \hat{\tilde{S}}_z\cos{\theta}-\hat{\tilde{S}}_x\sin{\theta}\\
\hat{S}_x &=& \hat{\tilde{S}}_x\cos{\theta}+\hat{\tilde{S}}_z\sin{\theta}.
\label{spin_rotation}
\end{eqnarray}
Using Holstein-Primakoff (HP) transformation\cite{hp}, the spin operators along the classical spin vector can be represented by,
\begin{eqnarray}
& & \hat{\tilde{S}}_z = S-a^\dagger a \\
& & \hat{\tilde{S}}_{+}=\sqrt{(2S-a^\dagger a)}a ,\quad \hat{\tilde{S}}_{-}=a^\dagger\sqrt{(2S-a^\dagger a)}
\label{HP_spin}
\end{eqnarray}
where $\hat{\tilde{S}}_{\pm} = \hat{\tilde{S}}_x \pm \imath \hat{\tilde{S}}_{y}$. For large spin, the operators and the Hamiltonian can be expanded in $1/S$. After substituting Eq.\ref{HP_spin} and Eq.\ref{spin_rotation} in the Hamiltonian (Eq.\ref{spin_ham}), we eliminate the linear terms of the bath modes by the transformation $c_{k} = b_{k} + \frac{\lambda}{4\hbar \omega}\sqrt{2S}\cos{\theta}$. Finally the Hamiltonian can be written systematically in descending orders in total spin $S$,
\begin{equation}
H = S H_{0} + \sqrt{S} H_{1} + H_{2} + O(1/\sqrt{S})
\label{H_exp}
\end{equation}
where higher order terms in $1/S$ are neglected. The first term $H_0$ corresponds to the energy of the classical spin,
\begin{equation}
H_0 = -J\sin{\theta}+\frac{U}{2}\cos^2{\theta}- \frac{\gamma_{-}}{2}\cos{\theta} -\frac{\gamma_{+}}{8}\cos^2{\theta}.
\label{cl_en}
\end{equation}
Here last two terms in the classical energy are generated due to the coupling with the bosonic bath. Elimination of bath degrees of freedom generates an effective ferromagnetic interaction term in the classical Hamiltonian which competes with the anti-ferromagnetic interaction due to repulsive interaction $U$ and gives rise to a quantum phase transition.
In this section we consider the model in absence of asymmetry, $\gamma_{-} =0$, to understand symmetry breaking related to quantum phase transition and dynamical transition. The asymmetric coupling $\gamma_{-}$ acts as an additional magnetic field which changes the nature of the transition which is briefly discussed in next section.
Extremizing the classical energy gives possible solutions, $\cos\theta =0$ or $\sin\theta = -J/\tilde{U}$, where $\tilde{U}=U-\gamma_{+}/4$. 
When $\tilde{U}>0$, the spin interaction is anti-ferromagnetic and $\cos\theta =0$ corresponds to the ground state with no net magnetization along the $z$-axis. However $\sin\theta = J/\tilde{U}$ represents a dynamical steady state for $\tilde{U}>J$ known as self-trapped state\cite{BJJ_osc} which will be discussed in the next section. 
On the other hand for strong coupling with the bath, when $\gamma_{+}>4 U$, ferromagnetic interaction dominates and $\sin\theta = J/|\tilde{U}|$ describes the symmetry broken ground state with non vanishing magnetization $\cos\theta = \pm \sqrt{1 -(J/|\tilde{U}|)^2}$  (equivalently a number imbalance  between two condensates). The bifurcation in imbalance fraction $z= (n_{L} -n_{R})/N = \cos\theta $ as shown in Fig.\ref{fig:1}(a) represents the quantum phase transition and the system chooses one of the branches with $z\neq 0$ above the transition.

The term $H_1$ proportional to $\sqrt{S}$ in the Hamiltonian vanishes for the above values of $\theta$ corresponding to the ground state and the steady states. The last term $H_2$ describes the fluctuations around the ground state which can be written as,
\begin{eqnarray}
H_2 = && \epsilon a^\dagger a + \Delta(a^2+a^{\dagger 2})+\sum_k\tilde{\lambda}_k(a+a^\dagger)(c_k+c^{\dagger}_k)\nonumber\\
&& +\sum_k\hbar \omega_k c^{\dagger}_k c_k + \Delta,
\label{ham2}
\end{eqnarray}
Where we define the constants in terms of $\theta$,
\begin{eqnarray}
\epsilon &=& \frac{J}{\sin{\theta}}+\frac{U}{2}\sin^2{\theta}\\
\Delta &=& \frac{U}{4}\sin^2{\theta}
\label{theta_delta} 
\end{eqnarray}
Above Hamiltonian describing the fluctuations is similar to the `Caldeira-Leggett' model of harmonic oscillator where the excitations of spin system are coupled to the effective bath modes. 
Since $H_2$ is quadratic in bosonic operators, it can be diagonalized by 
unitary transformations\cite{ambegaokar,solenov},
\begin{equation}
a_{\alpha}=\sum_\beta A_{\alpha\beta}q_\beta+\bar{A}_{\alpha\beta}q^\dagger_\beta
\label{a_modes}
\end{equation}
where we denote the original bosonic operators by $a_\alpha=(a,c_k)$, $A_{\alpha\beta}$, and $\bar{A}_{\alpha\beta}$ are elements of complex matrices ensuring the canonical commutation relations between new bosonic operators, $[q_{\beta},q^{\dagger}_{\beta'}]=\delta_{\beta,\beta'}$. In terms of these new operators $q_{\beta},q^{\dagger}_{\beta}$, the Hamiltonian can be written in diagonal form,
\begin{equation}
H_2=\sum_\beta \mathcal{E}_\beta q^\dagger_\beta q_\beta +\varepsilon_0,
\label{diag_H2}
\end{equation}
where the excitation energies are given by $\mathcal{E}_{\beta}$ and $\varepsilon_0$ is the correction to the ground state energy due to quantum fluctuation. 
We follow the standard procedure to obtain new operators and excitation spectrum as outlined in \cite{ambegaokar,solenov}. 
Using Eq.\ref{a_modes} and Eq.\ref{diag_H2}, we obtain,
\begin{equation}
\sum_\beta (-A_{\alpha\beta}\mathcal{E}_\beta q_\beta+\bar{A}_{\alpha\beta}\mathcal{E}_\beta q^\dagger_\beta)\ = [H_2,a_{\alpha}],
\label{eqn_A}
\end{equation}
where the right hand side commutator can be calculated from the original Hamiltonian given in Eq.\ref{ham2} and then it can be written in terms of $q_{\beta}$,$q^{\dagger}_{\beta}$ using Eq.\ref{a_modes}.
Finally equating the coefficients of the operators $q_{\beta}$ and $q^{\dagger}_{\beta}$, both sides of the Eq.\ref{eqn_A} yields following
set of equations,
\begin{subequations}
\begin{eqnarray}
&(\mathcal{E}_{\beta} -\epsilon_0)A_{0\beta} =2\Delta\bar{A}^{\star}_{0\beta}+\sum_k\tilde{\lambda}_k(A_{k\beta}+A^{\star}_{k\beta})\\
&(\mathcal{E}_{\beta}+\epsilon_0)\bar{A}_{0\beta}=-2\Delta A^{\star}_{0\beta} -\sum_k\tilde{\lambda}_k(\bar{A}_{k\beta}+A^{\star}_{k\beta})\\
&(\mathcal{E}_{\beta}-\hbar\omega_k)A_{k\beta}=\tilde{\lambda}_k(A_{0\beta}+\bar{A}^{\star}_{0\beta})
\label{eigen_sys1}\\
&(\mathcal{E}_{\beta}+\hbar\omega_k)\bar{A}_{k\beta}=-\tilde{\lambda}_k(\bar{A}_{0\beta}+A^{\star}_{0\beta}).
\label{eigen_sys}
\end{eqnarray}
\end{subequations}
Where we have defined $\tilde{\lambda_k}=-\lambda_k\sin{\theta}/4$. Solving these eigenvalue equations,the excitation energies can be obtained from,
\begin{equation}
\mathcal{E}^{2}_{\beta}=\epsilon^{2}_0-4\Delta^2+4(\epsilon_0-2\Delta)\sum_k\frac{\tilde{\lambda}^{2}_{k}\hbar\omega_k}{\mathcal{E}^{2}_{\beta}-\hbar^{2}\omega^{2}_k}
\label{eigen_energy}
\end{equation}
where the parameters $\epsilon$ and $\Delta$ depends on the steady states obtained from the minimization of classical energy $H_0$.
Corresponding to a given eigenvalue, the transformation matrix elements are given by,
\begin{subequations}
\begin{eqnarray}
&& A_{0\beta}=(\mathcal{E}_{\beta}+\epsilon_0-2\Delta)\mathcal{N}_{\beta}\\
&& \bar{A}_{0\beta}=-(\mathcal{E}_{\beta}-\epsilon_0 + 2\Delta)\mathcal{N}_{\beta}
\label{A0}
\end{eqnarray}
\end{subequations}
Where the normalization constant is,
\begin{equation}
\mathcal{N}_\beta=\frac{1}{\sqrt{4(\epsilon_0-2\Delta)\mathcal{E}_{\beta}}}\bigg(1+\sum_k 4\hbar\omega_k\tilde{\lambda}^{2}_{k}\frac{(\epsilon_0-2\Delta)}{(\mathcal{E}^{2}_{\beta}-\hbar^2\omega^{2}_k)^2}\bigg)^{-\frac{1}{2}}.
\label{N_beta}
\end{equation}
Other elements $A_{k\beta}$ and $\bar{A}_{k\beta}$ can be obtained from the relation Eq.\ref{eigen_sys1},Eq.\ref{eigen_sys}.
For weak coupling to the bath modes, the classical  spin is aligned along the $x$ axis and the ground state corresponds to $\sin\theta = 1$.
By increasing the coupling strength $\gamma_{+}$, it is evident from Eq.\ref{eigen_energy} that the lowest excitation energy vanishes at a critical coupling strength,
\begin{equation}
\gamma_{+c}/4 = J + U
\label{crit_gamma}
\end{equation} 
which signifies the instability of the ground state. Further increasing the coupling $\gamma_{+}$, the system chooses a new symmetry broken ground state described by $\sin\theta = J/(\gamma_{+}/4-U)$ and appearance of energy gap ensures its stability(see Fig.\ref{fig:1}(c) and Fig.\ref{fig:1}(d)). From the minimization of classical energy and vanishing of energy gap as shown in Fig.\ref{fig:1}(c) and Fig.\ref{fig:1}(d), it is evident that the BJJ undergoes a quantum phase transition at a critical coupling with the bosonic bath given in Eq.\ref{crit_gamma}. This unique phenomena in BJJ and appearance of ground state with number imbalance results from coupling with the bosonic bath modes. In spin-boson model, a similar ferromagnetic transition occurs by coupling a single spin with bosonic bath modes, however nature of such transition crucially depends on the 
spectral properties of the bath\cite{bulla}. In absence of any asymmetry,  $\gamma_{-}=0$, BJJ undergoes a continuous transition which occurs even in the presence of a single bosonic mode as shown in Fig.\ref{fig:1}(c). In the absence of bath degrees, substituting $\lambda_k =0$ in Eq.\ref{eigen_energy}, we recover the Josephson oscillation frequency\cite{BJJ_osc},
\begin{equation}
\omega_{JJ} =\mathcal{E}_{0}/\hbar = \sqrt{J(J+U)}/\hbar.
\label{jj_freq}
\end{equation} 

Within HP method, we can also calculate the quantum fluctuations in the vicinity of the critical point. For the ground state satisfying the relation $q_{\beta}|0\rangle = 0$, we can obtain the 
quantum correction to the classical energy $\varepsilon = \langle 0|H_2|0\rangle$. Near the quantum phase transition, energy due to quantum fluctuation diverges as $\varepsilon \sim 1/\sqrt{\Delta_{E}}$, where $\Delta_{E}$ is the energy gap. Also the spin fluctuation $\chi = \langle \hat{S}_{z}^{2}\rangle -\langle \hat{S}_{z}\rangle ^{2}$ related to the fluctuation of number imbalance in BJJ can be calculated from the expression,
\begin{eqnarray}
\chi =&& \frac{S}{2}\sin^2{\theta}+S\sin^2{\theta}(\sum_\beta|\bar{A}_{0\beta}|^2+\sum_\beta A_{0\beta}\bar{A}_{0\beta})\nonumber\\
&&-(\sum_\beta|\bar{A}_{0\beta}|^2)^2\cos^2{\theta}
\label{spin_fluc}
\end{eqnarray}
Which also diverges as $\sim 1/\sqrt{\Delta_{E}}$ near the quantum critical point as shown in Fig.\ref{fig:1}(b).
In the next section, we consider the out of equilibrium dynamics of Josephson oscillation in presence of bath degrees.

\section{Dynamical equations and steady states} 
In this section, we study the dynamics of BJJ and dissipative effects induced by coupling it to the bosonic bath. The collective dynamics of a BJJ can be well described by time dependent variational principal\cite{df}. Since the BJJ is described by a spin Hamiltonian of a large spin $S\gg 1$, a spin coherent state is the natural choice for the variational wavefunction to study its dynamics. For BJJ coupled to the bosonic modes described by the Hamiltonian Eq.\ref{spin_ham}, we consider following time dependent variational wavefunction,
\begin{equation}
|\Psi(t)\rangle = |\mu(t)\rangle \prod_{k}|\alpha_{k}(t)\rangle
\label{var}
\end{equation}
where $|\mu \rangle$ is the spin coherent state and $|\alpha_{k}\rangle$ is coherent state of the bosonic bath mode with momentum index $k$, 
\begin{eqnarray}
&& |\mu\rangle = (1 + |\mu|^{2})^{-S} \exp{(\mu\hat{S}_{-})}|S_z =S\rangle,\\
&& |\alpha_{k}\rangle = \frac{1}{\sqrt{\pi}}\exp{(-\frac{1}{2}|\alpha_{k}|^{2})}\exp{(\alpha_{k}\hat{c}^{\dagger}_{k})}|0\rangle.
\label{coherent_states}
\end{eqnarray}
The complex parameter of the spin coherent state can be written as $\mu = \tan(\theta/2)e^{\imath \phi}$ representing the orientation of the classical spin $\vec{S}=(S\sin\theta\cos\phi,S\sin\theta\sin\phi,S\cos\theta)$ \cite{radcliffe}. For coherent states of the bath modes, $\alpha_k$ have usual classical phase-space representation\cite{radcliffe}. Time dependent coherent state ansatz can also capture the dynamics of spin-boson system\cite{td_sb}.
In time dependent variational approach (TDVA)\cite{df}, the Lagrangian of the corresponding classical variables can be constructed from the relation,
\begin{eqnarray}
&&\mathcal{L}(\mu(t),\{\alpha_{k}(t)\}) =\langle \Psi(t)| \imath \hbar\frac{\partial}{\partial t} - \hat{H}|\Psi(t)\rangle \nonumber\\
&&=S\bigg[z\dot{\phi}+\sum_k(\dot{p}_{k}q_{k}-\dot{q}_{k}p_{k})+J\sqrt{1-z^2}\cos{\phi}-\frac{U}{2}z^2 \nonumber\\
&&-\frac{z}{2}\sum_k\lambda_{k}q_k-\frac{1}{2}\sum_k(p^2_{k}+q^2_{k})\hbar\omega_{k}+\frac{\gamma_{-}}{2}z\bigg]
\label{cl_lagrangian}
\end{eqnarray}
Where we have defined the canonical momentum $z=\cos{\theta}$ corresponding to the relative phase $\phi$ between the condensates in BJJ. Scaled dimensionless canonically conjugate variables of the oscillator modes are given as,
\begin{equation}
q_k=\frac{(\alpha_{k}+\alpha^{\star}_{k})}{\sqrt{2S}}, \quad p_k=-\imath\frac{(\alpha_{k}-\alpha^{\star}_{k})}{\sqrt{2S}}
\end{equation} 
From the Euler-Lagrange equation $\frac{d}{dt}(\frac{\partial \mathcal{L}}{\partial \dot{Q}})-\frac{\partial \mathcal{L}}{\partial Q}=0$ of the collective variables $Q= \{z,\phi,q_k,p_k\}$, one can arrive at the following set of equations of motion,
\begin{subequations}
\begin{eqnarray} \label{eqns of motion}
&&\dot{z}=-J\sqrt{1-z^2}\sin{\phi} \label{eqm1} \\
&&\dot{\phi}=J\frac{z}{\sqrt{1-z^2}}\cos{\phi}+Uz-\frac{\gamma_{-}}{2}+\sum_{k}\frac{\lambda_k}{2}q_k \label{eqm2}\\
&&\dot{q}_{k}=p_k\hbar\omega_{k} \label{eqm3}\\
&&\dot{p}_{k}=-\lambda_{k}\frac{z}{2}-q_k\hbar\omega_{k} \label{eqm4}
\end{eqnarray}
\end{subequations}
From the equation of motion of the classical collective coordinates, we study different oscillation modes of the BJJ and dissipative dynamics in the subsequent sections.

First we analyze the steady states and their stability of the dynamical equations. Steady states can be obtained by setting LHS of Eq.\ref{eqm1}-\ref{eqm4} to be zero and are denoted by $\{\bar{\phi},\bar{z}\}$. Here we classify the steady states according to the value of the relative phase $\bar{\phi}=0$ or $\bar{\phi}=\pi$. 
As mentioned earlier, for the ground state $\bar{\phi}=0$. Similarly the steady state value of imbalance $\bar{z}$ satisfies the following equation,
\begin{equation}
(J\cos\bar{\phi}+\sqrt{1-\bar{z}^2}\tilde{U})\bar{z}-\frac{\gamma_{-}}{2}\sqrt{1-\bar{z}^2}=0.
\label{steady_z}
\end{equation}
For $\gamma_{-} =0$, when $\tilde{U} = U -\gamma_{+}/4 < 0$, the ground state corresponding to $\bar{\phi}=0$ undergoes the quantum phase transition above $\gamma_{+c}= 4(U+J)$ for sufficiently strong coupling to the bath modes. On the other hand when $\tilde{U}>0$, the steady state with $\bar{\phi}=\pi$ exhibits a dynamical symmetry breaking above the critical interaction strength $\tilde{U}_{0} =J$. In both cases, the imbalance factor shows a bifurcation from $\bar{z}=0$ to $\bar{z} = \pm \sqrt{1 -J^2/|\tilde{U}|^2}$ as a result of symmetry breaking.  
It is important to note that for symmetry broken case with non vanishing imbalance (or magnetization), the bath oscillators are also shifted by an amount $\bar{q}_{k} = -\lambda_{k}\bar{z}/2\hbar\omega_{k}$. This phenomena is also associated with the phase transition in spin-boson system\cite{vondelft} and in the Dicke model\cite{brandes}.

Next we perform linear stability analysis of the steady state solutions and obtain the frequency of small amplitude oscillations. Around the steady states, the dynamical variables can be written as $Q(t)=\bar{Q} + \delta Q e^{\imath \omega t}$, and the eigen-frequencies $\omega$ of the  fluctuations $\delta Q$ can be obtained from the linearized equation of motion. The small amplitude oscillation frequencies around the steady states $\{\bar{\phi},\bar{z}\}$ can be obtained from the equation,
\begin{equation}
\omega^2=\frac{J^2}{(1-\bar{z}^2)}+\bigg(UJ+\frac{J}{4}\sum_k\frac{\lambda^2_{k}\omega_k}{\omega^2-\omega^2_{k}}\bigg)\sqrt{1-\bar{z}^2}\cos{\bar{\phi}}.
\label{ex_frequency}
\end{equation}
It is important to note that these small amplitude oscillation frequencies exactly match with the collective excitation frequencies $\mathcal{E}_{\beta}/\hbar$ in Eq.\ref{eigen_energy} obtained from HP approximation. This agreement indicates the correctness of the time dependent variational analysis used for large spin system.

In absence of bath modes, we recover the known Josephson oscillation frequencies for different steady states by substituting $\lambda_{k}=0$ in Eq.\ref{ex_frequency}. Quantum phase transition is absent for the ground state with $\{\bar{\phi}=0,\bar{z}=0\}$ and the normal ac Josephson oscillation has frequency $\omega_{JJ} = \sqrt{J(J+U)}/\hbar$. The oscillation around the steady state $\{\bar{\phi}=\pi,\bar{z}=0\}$ is known as Pi-oscillation mode which has frequency,
\begin{equation}
\omega_{\pi} = \sqrt{J(J-U)}/\hbar.
\label{pi_mode}
\end{equation}
This Pi-oscillation becomes unstable above a critical interaction strength $U_c =J$ and a symmetry broken `self-trapped' state appears with imbalance $\bar{z} \neq 0$. The oscillation frequency around the self-trapped state is given by,
\begin{equation}
\omega_{st} = U\sqrt{1-J^2/U^2}/\hbar.
\label{self_trapfreq}
\end{equation}
Above dynamical states, particularly the self-trapped state of BJJ arises due to interactions and have also been observed in experiments\cite{BJJ_expst}. In next section, we investigate the fate of such metastable dynamical states in presence of bosonic modes.

\section{BJJ coupled to a single mode} To gain more insight, we first analyze the dynamics of BJJ coupled to a single bosonic mode with frequency $\omega_{0}$ and coupling constant $\lambda$. In this case, the model in Eq.\ref{spin_ham} becomes a generalization of Dicke model\cite{dicke} with an additional interaction term. This model can describe the dynamics of BJJ in presence of cavity mode. 
The steady states and small amplitude oscillation frequencies can be obtained from Eq.\ref{steady_z} and Eq.\ref{ex_frequency}. As mentioned earlier, we can categorize the steady states by the relative phase $\bar{\phi}=0$ (ground state) and $\bar{\phi}=\pi$ (dynamical branch). 
For $\gamma_{-}=0$, in absence of any symmetry breaking field both the branches of steady states exhibits symmetry breaking corresponding to two different physical phenomena. Interestingly, even for single mode the ground state of BJJ undergoes quantum phase transition when $\gamma_{+}>\gamma_{+c}$ given in Eq.\ref{crit_gamma} and the symmetry breaking in imbalance fraction $\bar{z}$ is represented by the bifurcation of steady states as shown in Fig.\ref{fig:1}(a). Whereas, the other branch with $\bar{\phi}=\pi$ undergoes dynamical transition above a critical repulsive interaction strength $U_{c} = J + \gamma_{+}/4$, as discussed in previous section. 
Using Eq.\ref{ex_frequency}, the frequencies of small amplitude oscillations around the steady states with $\bar{\phi}=0,\pi$, can be calculated analytically. For the symmetry unbroken phase, the frequencies can be written as,
\begin{equation}
\omega^2=\frac{1}{2}\left[\omega^2_0+\omega^2_{su}\pm
\sqrt{(\omega^2_0-\omega^2_{su})^2+\omega^2_0J\gamma_{+}
\cos{\bar{\phi}}}\,\right]
\label{frequency_unbroken}
\end{equation}
where $\omega^2_{su}=J(J+U\cos{\bar{\phi}})$. For symmetry broken phase, the excitation frequencies are given by,
\begin{equation}
\omega^2=\frac{1}{2}\left[\omega^2_0+\omega^2_{sb}\pm \sqrt{(\omega^2_0-\omega^2_{sb})^2+\frac{\omega^2_0}{|\tilde{U}|}J^2\gamma_{+}\cos{\bar{\phi}}}\,\right]
\label{frequency_broken}
\end{equation}
where, $\omega^2_{sb}=\tilde{U}^2+U J^2\cos{\bar{\phi}}/|\tilde{U}|$.
For both the transitions, the lowest mode frequency vanishes as $\sim (|\tilde{U} -\tilde{U}_{c}|)^{1/2}$ at the critical point. 

\begin{figure}[ht]
\centering
\includegraphics[height=7cm,width=9cm]{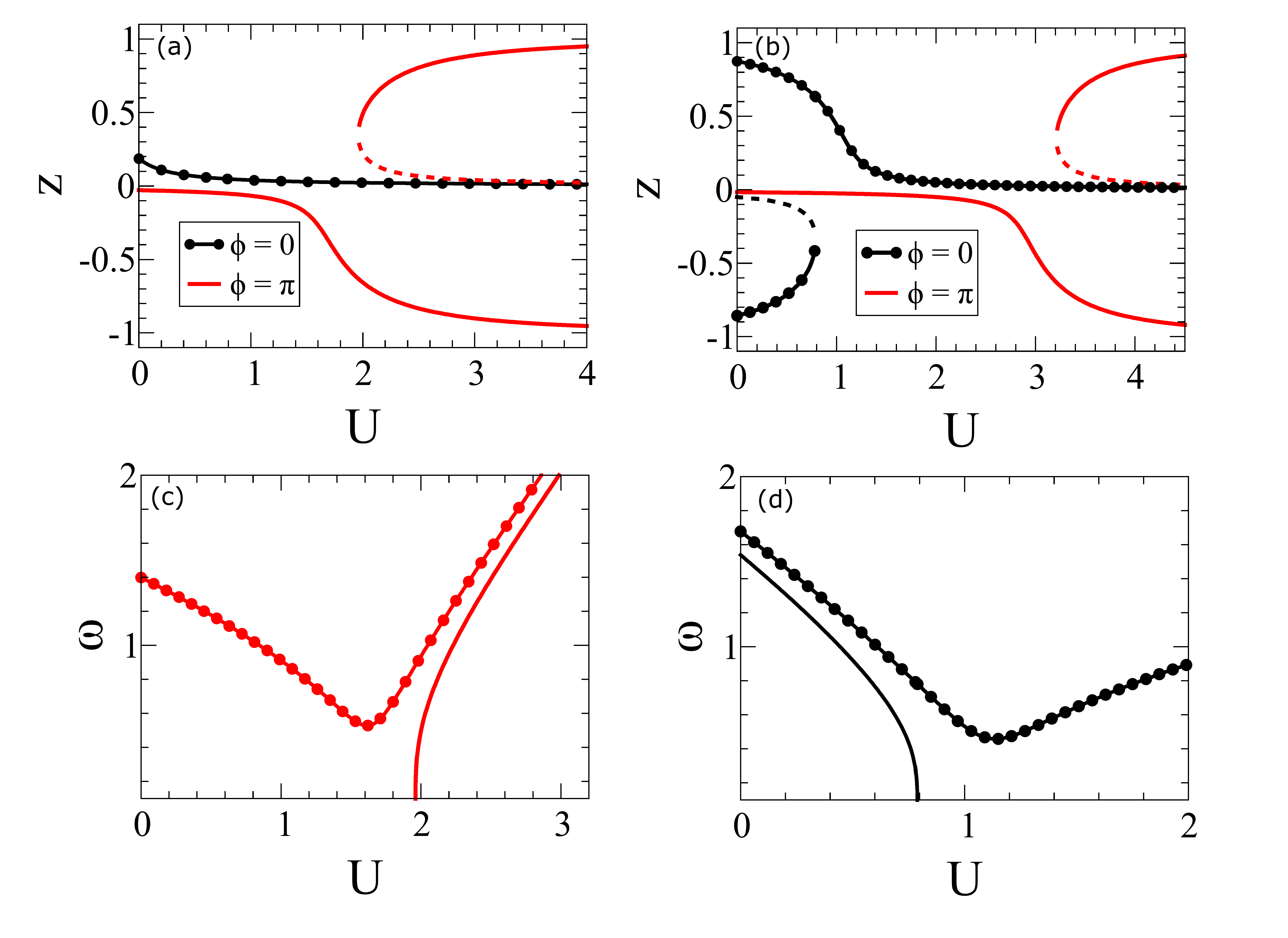}
\caption{Steady states and their stability in presence of weak asymmetry $\gamma_{-}=0.1$ and for single bath mode with $\omega_0=3.0$.
Stable (unstable) branches are denoted by solid (dotted) lines (a) for weak dissipative coupling $\gamma_{+}=0.75\gamma_{+c}$ and (b) for strong dissipative coupling $\gamma_{+}=2.0\gamma_{+c}$ showing QPT. (c) Lowest excitation energy corresponding to the stable steady state as shown in (a) for $\bar{\phi}=\pi$. (d) Same as (c) corresponding to stable steady state as shown in (b) for $\bar{\phi}=0$. In (c) and (d) solid dotted line represents continuous branches. For imperfect bifurcation, the gap does not vanish for continuous branches.
}
\label{fig:2}
\end{figure}

In presence of asymmetry $\gamma_{-} \ne 0$, the nature of transition changes and symmetry breaking is absent. Imperfect bifurcation of the steady state with imbalance fraction $\bar{z}$ is depicted in Fig.\ref{fig:2}(a) and (b) for different values of $\gamma_{+}$. Absence of imaginary part of the frequencies $\omega$ ensures dynamic stability of corresponding steady states.
For both types of transition, the lowest frequency corresponding to the continuous branch of steady state does not vanish, instead it reaches a minimum near the transition. On the other hand, the frequency corresponding to the stable isolated symmetry broken branch vanishes at the point where this steady state ends. This phenomena is similar to the spinodal point in first order transition. 

\begin{figure}[ht]
\centering
\includegraphics[height=4.25cm,width=8.5cm]{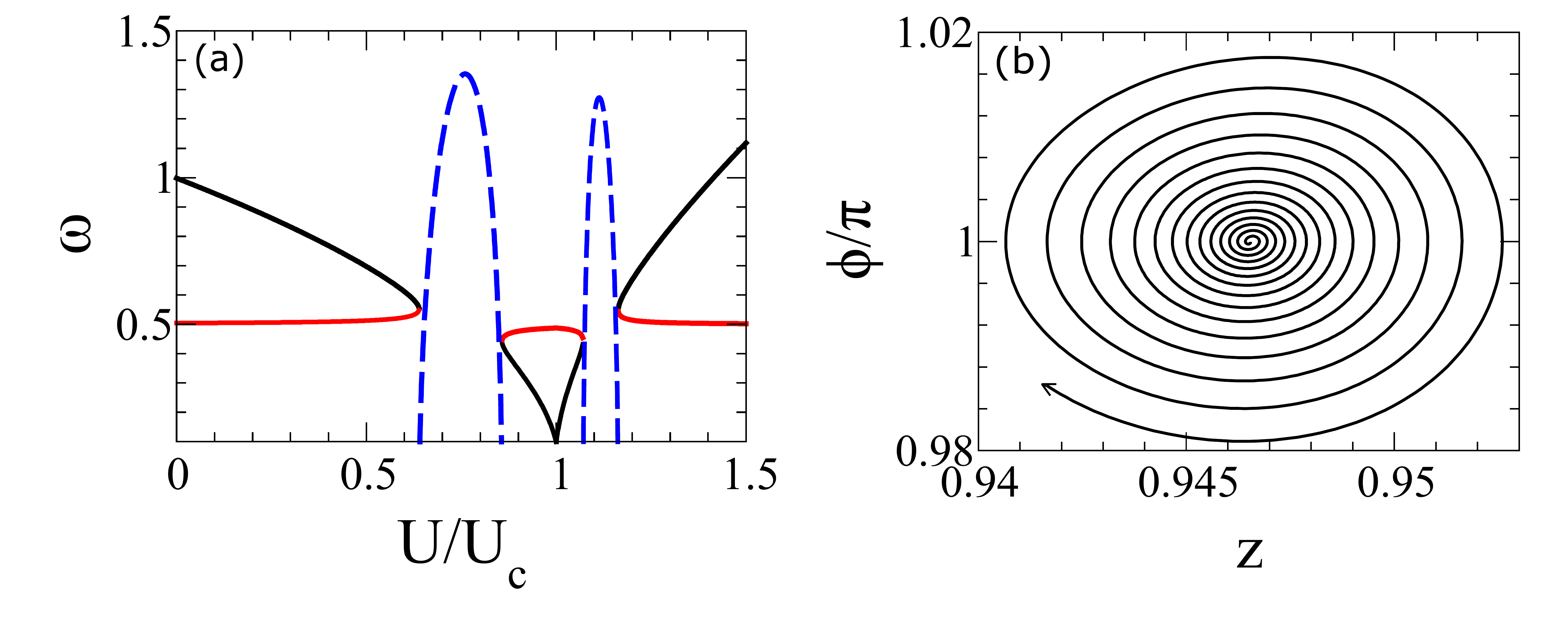}
\caption{(a) Oscillation frequencies $\omega$ corresponding to the steady state with $\bar{\phi}=\pi$ in presence of single mode with $\omega_0=0.5$ and $\gamma_+=0.01\gamma_{+c}$ showing dynamic phase transition at $U=U_c$. Real (imaginary) part of $\omega$ is denoted by solid (dashed) lines. In the plot imaginary part of $\omega$ is scaled by 10 times. (b) Phase space trajectory within the instability region for $U=2.91U_c$, $\omega_0=3.0$ and $\gamma_{+}=0.02\gamma_{+c}$.}
\label{fig:5}
\end{figure}

Interestingly, the dynamical steady with $\bar{\phi}=\pi$ becomes dynamically unstable within certain interval of interaction strength $U$ as shown in Fig.\ref{fig:5}(a). Physically this signify the dynamical instability of Pi-oscillation mode and self-trapped state when their frequency of oscillations are in resonance with the bath mode. 
As a consequence of this instability, the closed orbits near the fixed point becomes unstable and spiral out from the fixed point as seen from Fig.\ref{fig:5}(b). The Pi oscillation becomes unstable within the region $\omega^2_0-\omega_{0}\sqrt{J\gamma_{+}}<\omega^2_{su}<\omega^2_0+\omega_{0}\sqrt{J\gamma_{+}}$ and within the range $\omega^2_0-\omega_{0}\sqrt{J^2\gamma_{+}/|\tilde{U}|}<\omega^2_{sb}<\omega^2_0+\omega_{0}\sqrt{J^2\gamma_{+}/|\tilde{U}|}$ the self trapped state is dynamically unstable. This phenomena can be tested in experiment by coupling the BJJ with a cavity mode of appropriate frequency. 

\begin{figure}[ht]
\centering
\includegraphics[height=4.5cm,width=8.75cm]{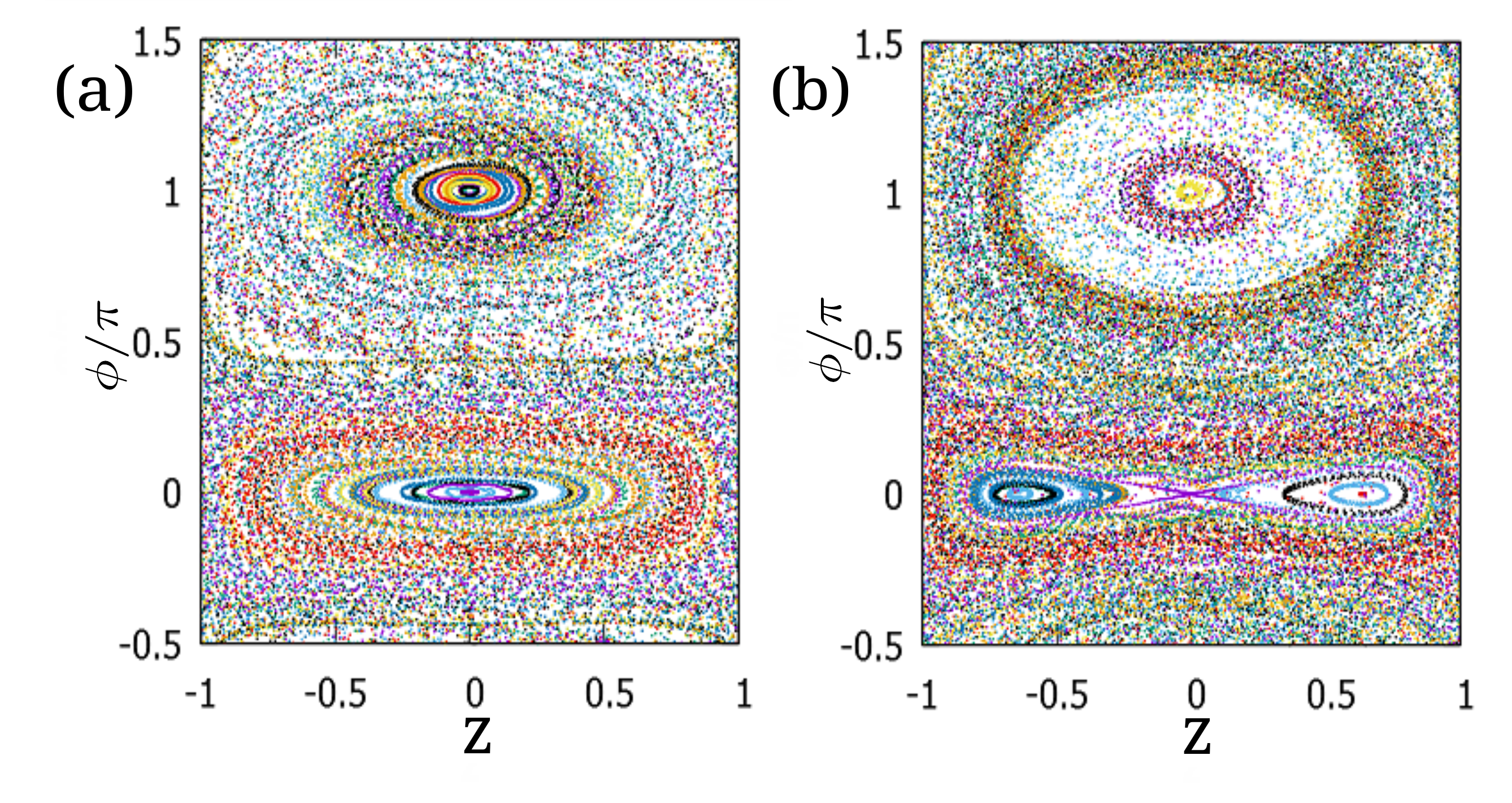}
\caption{Phase portraits showing QPT in presence of single bath mode. (a) Symmetry unbroken phase with $\gamma_+=0.75\gamma_{+c}$ and (b) symmetry broken phase with $\gamma_{+}=1.16\gamma_{+c}$. For both figures, $U=0.2$ and $\omega_0=3.0$.}
    \label{fig:3}
\end{figure}

To understand the dynamical behaviour, we analyze the phase-space trajectories in $\phi-z$ plane by solving full dynamical equations Eq.\ref{eqm1}-\ref{eqm4}. The phase space trajectories before and after the quantum phase transition are shown in Fig.\ref{fig:3}(a) and (b), where the closed orbits represents small amplitude oscillation around the stable fixed points. After the transition appearance of two fixed points represents possible symmetry broken ground states. The closed orbits at $\phi =\pi$ represents dynamically stable Pi oscillation.

\begin{figure}[ht]
\centering
\includegraphics[scale=0.18]{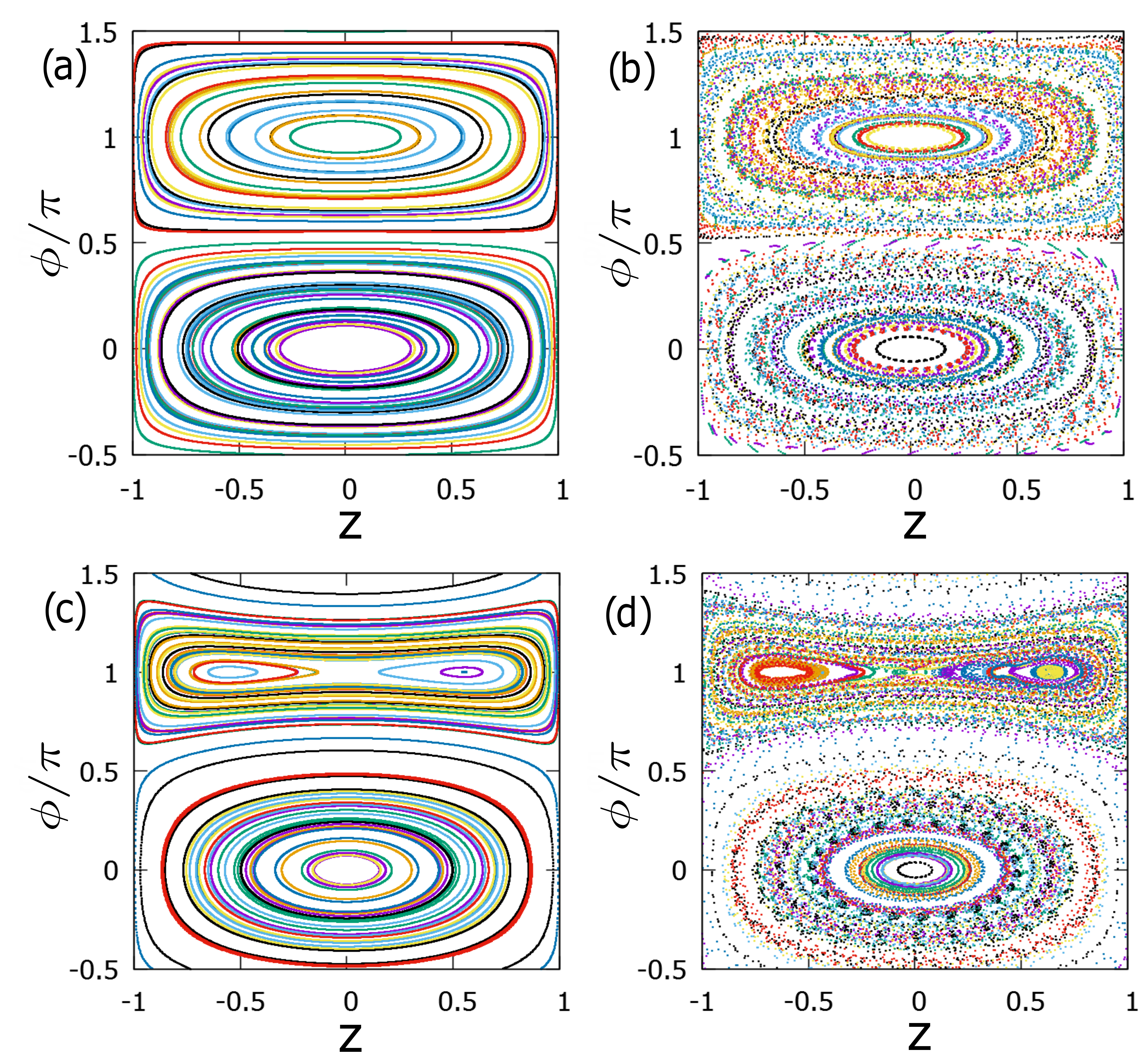}
\caption{Phase portraits showing dynamic phase transition in BJJ without(a,c) and with(b,d) single bath mode. (a) Symmetry unbroken phase with $U = 0.2U_c$ and (c) symmetry broken phase with $U = 1.2U_c$ in absence of bath mode. (b) Symmetry unbroken phase with $U = 0.41U_c$ and (d) symmetry broken phase with $U=1.25U_c$ in presence of single mode with $\gamma_+ = 0.2$, $\omega_0=3.0$ and $\Delta \epsilon=0.1$. }
\label{fig:4}
\end{figure}

To investigate the fluctuations induced by the bath mode in the dynamics of BJJ, the initial coordinates of the bath mode $\{p,q\}$ are randomly chosen from a Gaussian distribution with energy fluctuation $\Delta\epsilon$ around the fixed points. When the phase portraits of BJJ coupled to a single mode is compared with that of an isolated BJJ (see Fig.\ref{fig:4}), a diffusive behaviour in the phase space trajectories is observed due to the fluctuation of the bath modes. Near the quantum phase transition when BJJ is strongly coupled to the bosonic mode, the regular phase space trajectories around the fixed points shrinks and most of the region of the phase portrait in Fig.\ref{fig:3} exhibits chaotic behaviour. Such chaotic behaviour has also been observed in the dynamics of Dicke model indicating thermalization\cite{altland}. Ergodicity in BJJ has also been investigated in\cite{odell}. However, ergodicity and thermalization in this model is beyond the scope of present discussion and requires more detailed analysis. 
We expect high entanglement near the quantum phase transition may lead to chaotic dynamics and incoherent Josephson oscillation which is reflected in the phase portrait. Loss of phase coherence due to the bath modes is investigated in next section.

\section{BJJ coupled to Ohmic bath} In this section we discuss excitation spectrum and dissipative dynamics of Bose Josephson junction coupled to a bath containing many bosonic modes. Typically a bath is modeled by bosons with continuous energy spectrum up to certain cutoff energy larger than the energy scale of the system. The coupling strength, spectral density and temperature of the bath play crucial role in dissipative dynamics of the system coupled to it. The spectral density of the bath is defined as,
\begin{equation}
\mathcal{J}(\nu) = \sum_{k} \lambda_{k}^{2} \delta(\nu - \omega_{k}).
\label{spec_den}
\end{equation}
Here we consider Ohmic bath spectral density,
\begin{equation}
\mathcal{J}(\nu) = \frac{\alpha' \nu \nu_{c}^2}{\nu^{2} + \nu_{c}^{2}}\theta(\Omega - \nu),
\label{ohmic}
\end{equation}
where $\theta(x)$ is Heaviside step function, $\alpha'$ is the coupling constant and $\Omega$ is the cutoff frequency of the bath. Here $\nu_c$ is a parameter related to the memory effect. For $\nu \ll \Omega,\,\nu_{c}$, the spectral density follows $ \mathcal{J}(\nu) \sim \alpha' \nu$ corresponding to Ohmic bath\cite{weiss}. For the purpose of numerical calculations, we consider $N_{s}$ number of equispaced bath modes up to the cutoff frequency $\Omega$, so that the bath frequencies and coupling constants are given by,
\begin{equation}
\omega_{n}=\Delta n;~~\lambda_{n}^{2} = \mathcal{J}(\omega_{n})\Delta,
\label{discretization}
\end{equation}
where $n=0,1...N_{s}$ and $\Delta = \Omega/N_{s}$ is the spacing between energy levels.
According to this discretization, the effective coupling constant for dissipation can be written in terms of the spectral density as,
\begin{equation}
\gamma_{+} = \sum_{n}\frac{\lambda_{n}^{2}}{\hbar \omega_{n}} = \int_{0}^{\Omega}\frac{\mathcal{J}(\nu)}{\hbar \nu}d\nu
\label{gamma_con}
\end{equation}
which reduces to $\gamma_{+} = \alpha' \nu_{c}/\hbar \tan^{-1}\frac{\Omega}{\nu_{c}}$ for the spectral density given in Eq.\ref{ohmic}. 

\begin{figure}[H]
\centering
\includegraphics[height=5.75cm,width=7.75cm]{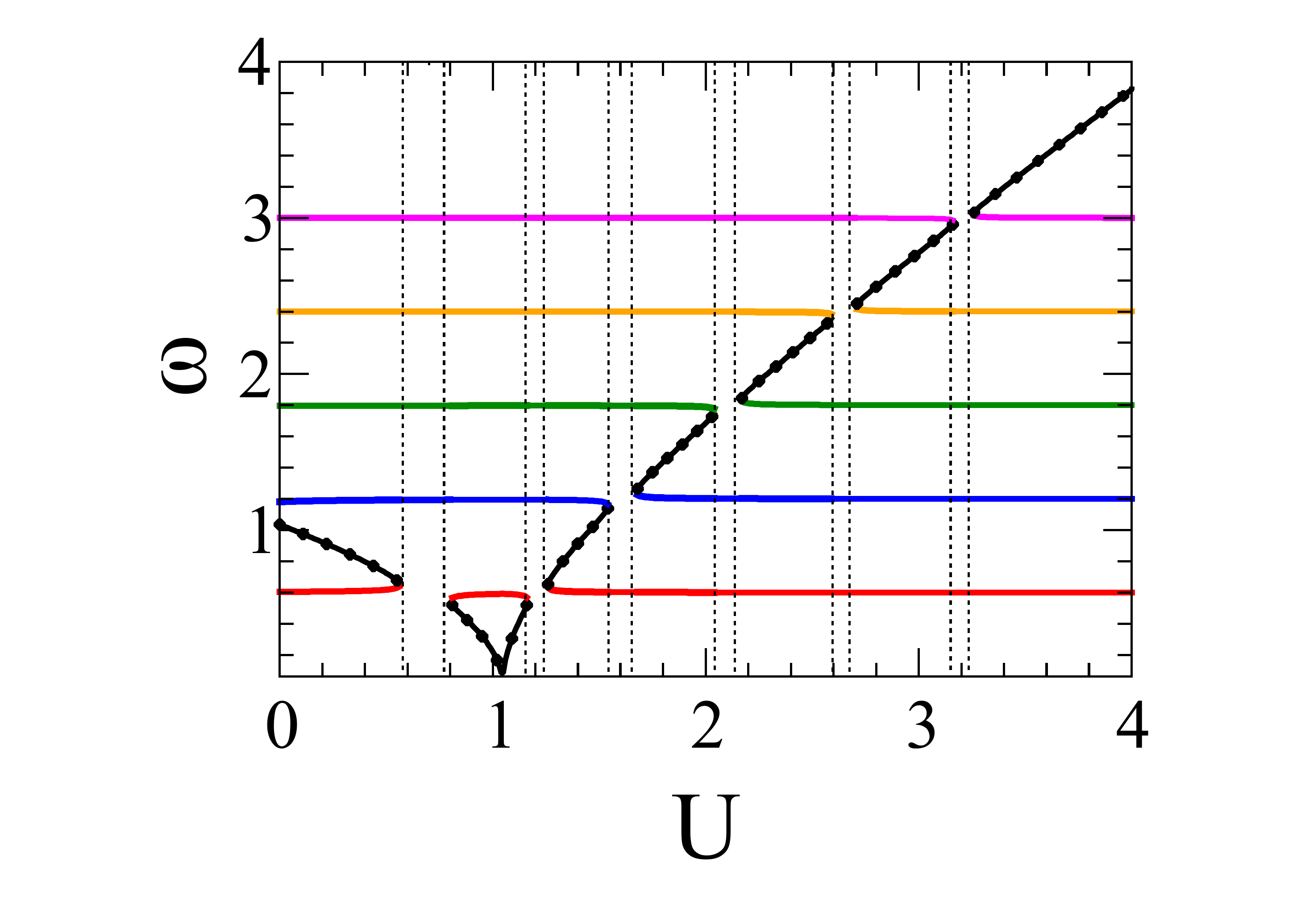}
\caption{Excitation frequencies corresponding to the steady state with $\bar{\phi}=\pi$ in presence of five bath modes exhibiting dynamic phase transition at $U=U_c=1.04$ where energy gap vanishes. Dynamically unstable regions of the steady state are demarcated by vertical dotted lines. Bath modes are chosen according to Eq.\ref{discretization}, with
$\Omega=3$ and $\alpha=0.08$.}
\label{fig:6}
\end{figure}

First we investigate the frequency of Josephson oscillation in the presence of few discrete bath modes for two types of steady states discussed in section III. The excitation spectrum of the ground state with relative phase $\bar{\phi}=0$ is shown in Fig.\ref{fig:1}(d). The lowest energy mode vanishes at the quantum phase transition. Bath modes remains unaffected except near the avoided level crossing. For this branch of steady state all frequencies are real, which ensures the stability of the ground state.
We compute the excitation spectrum corresponding to the steady state with $\bar{\phi}=\pi$. In this case, the lowest branch of excitation also vanishes at $\tilde{U}_c$ signifying the dynamical transition between Pi-oscillation and self-trapped state with $\bar{z} \neq 0$. However as observed for single mode, the excitation frequencies are not real always and pick up an imaginary part in the region where the crossing of energy levels occur as indicated by vertical dotted lines in Fig.\ref{fig:6}. 
Appearance of complex frequencies indicate dynamical instability of Pi-oscillation and self-trapped states in these parameter regions. This is an interesting effect arising due to the coupling with the bath modes which can be detectable in experiments.
\begin{figure}[H]
\centering
\includegraphics[height=7.5cm,width=9cm]{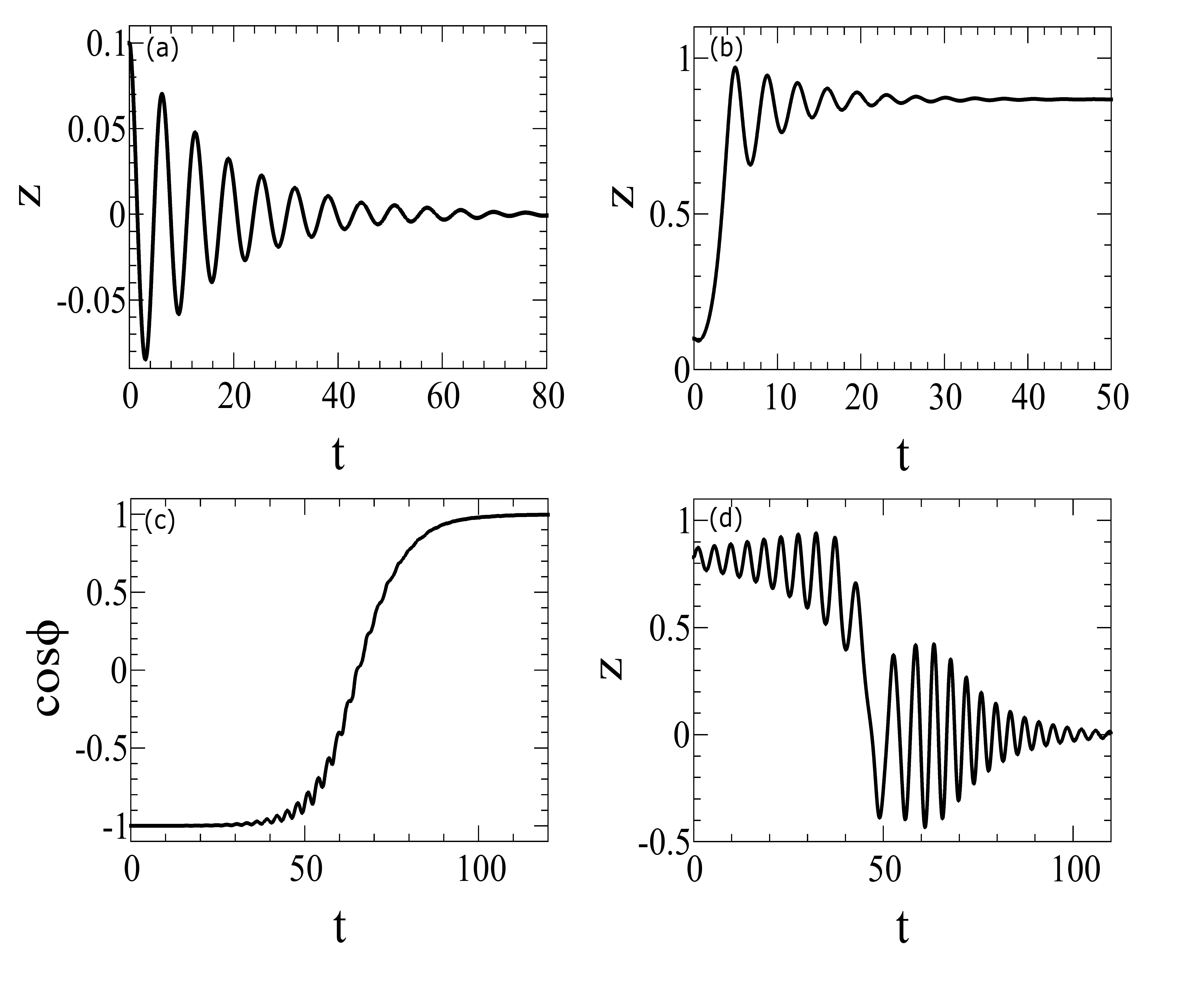}
\caption{Dynamics of BJJ in presence of Ohmic bath. 
Time evolution of number imbalance fraction $z$ for $U=0.5$ corresponding to (a) normal Josephson oscillation for weak dissipative coupling $\alpha=0.35\alpha_c$ and (b) symmetry broken phase with $\alpha=1.67\alpha_c$.
Decay from (c) steady state corresponding to Pi-oscillation for $U=0.31U_c$
and (d) self-trapped state for $U=1.45U_c$.
Parameters used for numerical simulation with $N_{s}=200$ bath modes are
$\Omega = 10$, $\nu_c = 8$ and $\Delta \epsilon = 0.1$.
}
\label{fig:7}
\end{figure}

Next, we study the fate of different types of Josephson oscillations in the presence of dissipation arising from a bath of quasi-continuous modes. The dynamical equations Eq.\ref{eqm1}-\ref{eqm4} are time evolved from a suitably chosen initial condition for large number of bath modes with spectral density given in Eq.\ref{discretization}. Typical parameters chosen for simulation are $N_{s}=160$, $\hbar \Omega/J = 10$ and $\hbar\omega_c/J=1$. Initial coordinates of the bath oscillators $\{p_{i},q_{i}\}$ are chosen randomly from a Gaussian distribution around an appropriate steady state value so that each bath mode has average energy $\Delta\epsilon$ due to the fluctuations. In this way we can incorporate quantum as well as thermal fluctuations in the dynamics\cite{wigner_spinboson}, however away from the quantum critical point, the quantum fluctuation is small for large spin system (equivalently with large number of bosons), since $\hbar_{eff} \sim 1/S$.   
The time evolution of the dynamical variables of the BJJ $z(t)$ and $\phi(t)$ are obtained after ensemble averaging of $\mathcal{N}_{en} \sim 10^3$ number of initial realization of the bath variables describing equilibrium state of the bath. To avoid the revival due to discretization of bath frequency, the time evolution is restricted up to a time scale $\leq 2\pi/\Delta$.

The dynamics of imbalance fraction around the steady state $\bar{z}=0,\bar{\phi}=0$ in presence of weak dissipation is shown in Fig.\ref{fig:7}(a). Dissipative environment gives rise to damping in Josephson oscillations and the imbalance fraction eventually relaxes to its steady value $z=0$. It is known that coupling to Ohmic bath introduces damping $\gamma_{D} \sim \alpha'$ in the Markovian limit\cite{cl}. On the other hand, in the strong coupling regime with $\gamma_{+}>\gamma_{+c}$, the imbalance fraction quickly saturates to symmetry broken steady state  signifying the quantum phase transition (see Fig.\ref{fig:7}(b)).

Both the Pi-oscillations and self-trapped states are very unique in nature since they arise dynamically in an isolated BJJ as a result of inter atomic interactions. Fate of such metastable states in the presence of dissipation is a pertinent issue. For Pi-oscillation, we consider the initial state of the Josephson junction close to the steady state with $\bar{z}=0$,$\bar{\phi}=\pi$ and study the dynamical evolution of the relative phase $\phi$ under the influence of bath modes. In dissipative environment, this metastable state becomes unstable and decays to the ground state with $\phi=0$, as a result the value of $\cos\phi$ changes from -1 to 1 as shown in Fig\ref{fig:7}(c). Similarly the imbalance fraction $z$ of an initially prepared self-trapped state with $z\neq 0$ also vanishes in dynamical evolution indicating relaxation to the ground state (as shown in Fig.\ref{fig:7}(d)). Both the Pi-oscillation and self-trapped state corresponding to the metastable steady state with $\bar{\phi}=\pi$ do not survive under dissipative dynamics and relax to the ground state by exchanging energy with the bosonic bath.

\begin{figure}[H]
\centering
\includegraphics[height=4cm,width=9cm]{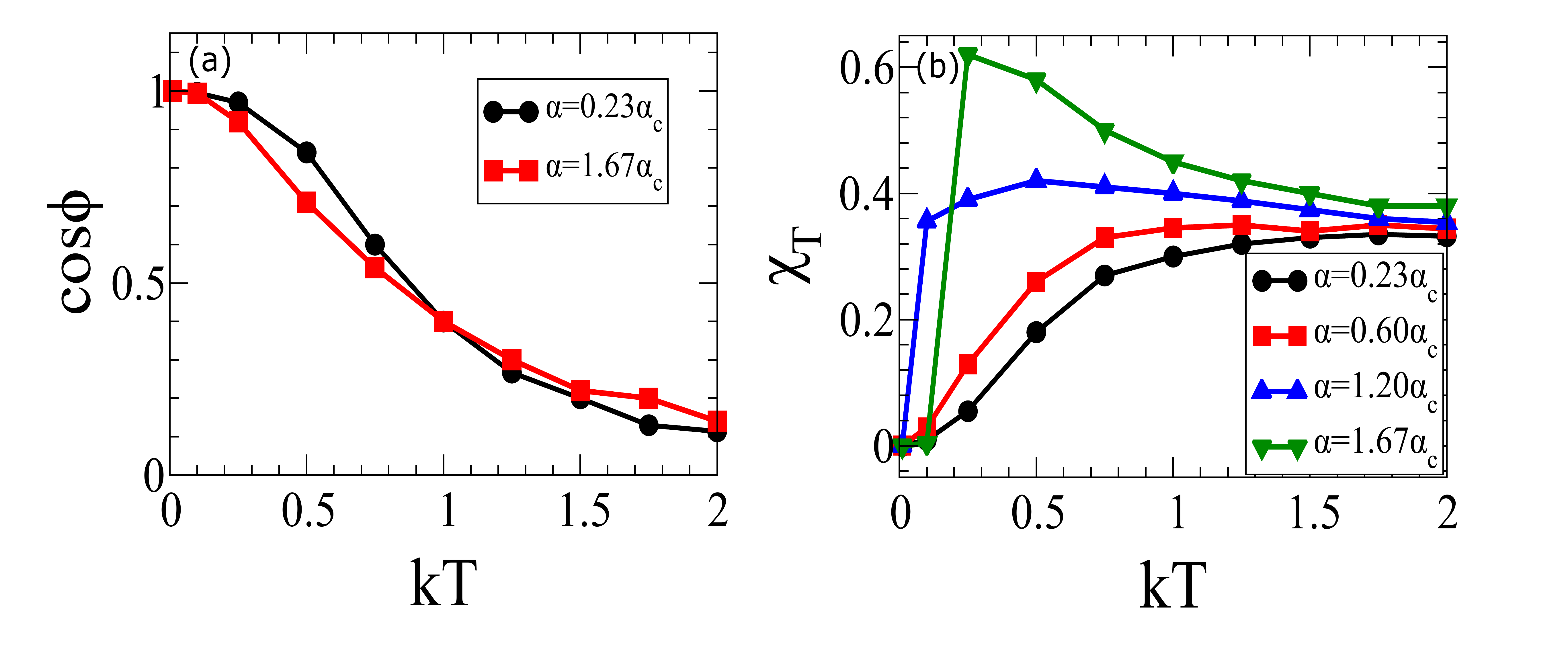}
\caption{(a) Phase fluctuation and (b) fluctuation of imbalance fraction as a function of temperature of the Ohmic bath for different values of $\alpha$. Parameters used are $U=0.5<U_c$, $\Omega = 10$, $\nu_{c} = 8$ and $N_{s}=200$.
}
\label{fig:8}
\end{figure}

Apart from dissipative dynamics and relaxation phenomena, fluctuation phenomena related to the loss of coherence in Josephson dynamics is another important effect arising due to the presence of heat bath. The relative phase $\phi$ between the condensates in two sites of the BJJ is the relevant dynamical variable describing the coherent Josephson oscillation. Fluctuation in relative phase $\phi$ is introduced by the dissipative environment and expected to increase with temperature leading to a loss of coherence.  This phase diffusion phenomena can be quantified from the average value of $\cos\phi$ since $\langle \cos\phi\rangle \approx e^{-\langle \phi\rangle^{2}/2}$. Here we study the phase fluctuation in the BJJ coupled to the bosonic bath from ensemble averaging of $\cos\phi$ after the system is evolved sufficiently long time. To incorporate thermal effect, we consider the bosonic bath is in equilibrium at temperature $T$, so that the average energy fluctuation of each classical bath mode $\{p_{i},q_{i}\}$ follows equipartition $\Delta\epsilon = k_{B}T$. In Fig.\ref{fig:8}(a), temperature dependence of phase fluctuation is shown in terms of $\langle \cos\phi\rangle$. With increasing temperature phase fluctuation increases, as a result $ \langle \cos\phi\rangle$ decreases from one. This result captures the essential features of the experimentally observed temperature dependence of the phase fluctuation\cite{exp_phase}. However in the present model, the phase fluctuation also depends on the coupling strength $\alpha'$ of the heat bath. For the symmetry broken ground state with strong dissipative coupling $\gamma_{+}>\gamma_{+c}$, the phase fluctuation also increases as depicted in Fig.\ref{fig:8}(a), although the variation with the coupling strength is not very significant. In a similar manner, we calculate the thermal fluctuation of imbalance fraction $\chi_{T}=\langle\cos^{2}\theta\rangle -\langle\cos\theta\rangle^{2}$ shown in Fig.\ref{fig:8}(b). Above the critical coupling $\gamma_{+c}$, the imbalance fluctuation $\chi_{T}$ exhibits a peak which is reminiscence of quantum phase transition. 

\section{Conclusions} 
In this work, we study the effect of dissipation on Bose Josephson junction with two sites connected to two bosonic baths. The BJJ can effectively be described by a large spin system with anti-ferromagnetic interaction coupled to a single bath. A ferromagnetic interaction generated due to the bosonic modes leads to a quantum phase transition above a critical dissipative coupling. In the symmetry broken phase, the BJJ has nonzero number imbalance between the two sites. To analyze the quantum phase transition, the excitation spectrum and spin fluctuations (equivalently fluctuation in number imbalance) are calculated within Holstein-Primakoff approximation. At the critical point, the energy gap vanishes and quantum fluctuations show divergent behaviour. We analyze dynamics of the system within TDVA in terms of the collective coordinates of the coherent states. From the dynamical equations, we obtained the ground state as well as dynamical steady states related to Pi-oscillation and self-trapped state. We perform linear stability analysis to obtain the frequencies of small amplitude oscillation around the steady states. As a special case, the dynamics of BJJ coupled to a single bosonic mode is analyzed in details. Unlike the spin-boson model, in this case even a single mode can lead to the quantum phase transition with vanishing of energy gap. Interestingly, the dynamical steady state becomes unstable near the region where the frequency of oscillation is in resonance with the bath mode leading to a dynamical instability of self-trapped state and Pi-oscillation. 
This phenomena also persists for bath containing multi-mode and more unstable regions in the dynamical steady state appear. We also investigate the phase space trajectories of Josephson dynamics including small fluctuations in the bath mode. Appearance of chaotic regions in phase portrait in the vicinity of quantum phase transition indicates incoherent Josephson oscillation and entropy generation due to strong coupling with the bath mode. Such complex dynamics can be observed experimentally by coupling the BJJ with one or few cavity modes. 

For BJJ coupled to realistic Ohmic bath with large number of modes, we perform numerical simulation to study Josephson dynamics incorporating thermal fluctuations in the bath modes. For weak coupling with the bath, the Josephson oscillation of the imbalance fraction shows damping and relax to the ground state as a result of dissipation. On the other hand, above the critical coupling with the bath, the number imbalance quickly saturates to a non vanishing value corresponding to the new symmetry broken ground state. This is a clear manifestation of symmetry breaking controlled by dissipation.
The dynamical steady states corresponding to the Pi-oscillation and self-trapped states become unstable when BJJ is coupled to the bosonic bath and during time evolution they eventually relax to the ground state by exchanging energy with the bath modes. This is a significant effect of dissipation which can be tested in experiment. Dissipative environment plays a crucial role in phase coherent Josephson oscillation since the fluctuation of relative phase is enhanced by the bath modes. 
The temperature dependence of phase fluctuation is calculated in the presence of thermal bath which captures the experimentally observed behaviour. We also calculate temperature dependence of imbalance fluctuation which exhibits signature of quantum phase transition above the critical coupling. 

In summary, we studied effects arising from dissipation in a Bose Josephson junction coupled to a bosonic bath; among them most spectacular effect is the quantum phase transition to a state with finite number imbalance. In addition, the change in oscillation frequencies, relaxation dynamics, instability of Pi oscillation and self-trapped states and fluctuation in number imbalance are experimentally observable effects of dissipation. In experiments, BJJ has been realized by creating a condensate in double well trap and bosonic bath can be engineered by coupling the condensate with thermal Bose gas or by coupling it to another condensate where the phonon excitations can play the role of bosonic modes.

\section*{Acknowledgement}
We thank S. Ray for stimulating discussions.

\end{document}